\documentclass[12pt]{article} 
\usepackage{epsfig}

\begin {document}
%~\hfill ADP-AT-00-?????\\
%~\hfill  astro-ph/?????\\
~\hfill Astroparticle Physics, submitted\\[2cm]

\begin{center}
{\Large \bf Contribution of nuclei accelerated by gamma-ray 
pulsars to cosmic rays in the Galaxy}\\[1cm]
W. Bednarek$^1$ and R. J. Protheroe$^2$\\
$^1$Department of Experimental Physics, University of \L\'od\'z,\\
90-236 \L\'od\'z,  ul. Pomorska 149/153, Poland. \\ 
$^2$Department of Physics and Mathematical Physics,\\
The University of Adelaide, Adelaide, Australia 5005.\\
\end{center} 

\begin{abstract}
We consider the galactic population of gamma-ray pulsars as
possible sources of cosmic rays at and just above the ``knee'' in
the observed cosmic ray spectrum at $10^{15}$--$10^{16}$~eV. We
suggest that iron nuclei may be accelerated in the outer gaps of
pulsars, and then suffer partial photo-disintegration in the
non-thermal radiation fields of the outer gaps. As a result,
protons, neutrons, and surviving heavier nuclei are injected into
the expanding supernova remnant. We compute the spectra of nuclei
escaping from supernova remnants into the interstellar medium,
taking into account the observed population of radio pulsars. 

Our calculations, which include a realistic model for
acceleration and propagation of nuclei in pulsar magnetospheres
and supernova remnants, predict that heavy nuclei accelerated
directly by gamma-ray pulsars could contribute about 20\% of
the observed cosmic rays in the knee region.  Such a contribution
of heavy nuclei to the cosmic ray spectrum at the knee can
significantly increase the average value of $\langle\ln A\rangle$
with increasing energy as is suggested by recent observations.\\
\end{abstract}

\noindent PACS Numbers: 98.70.Sa (Cosmic rays), 97.60.Gb  (Pulsars).

\noindent Keywords: cosmic rays, spectrum and mass composition, 
pulsars, cosmic ray acceleration

\section{Introduction}

The cosmic ray spectrum is usually described by a superposition
of power laws with index $\sim 2.7$ below and $\sim 3.2$ above
the so-called ``knee'' at a few $10^{15}$eV. In the region of
$10^{18} - 10^{19}$ eV the spectrum flattens, and has an index
again close to $2.7$.  There are also suggestions that the shape
of the spectrum in the knee region has a fine structure which may
reflect acceleration of specific groups of nuclei, e.g helium,
oxygen, and iron, prominent in the Galactic abundances (see
\cite{ew98,ew99} and references therein).  Recent experimental
results show that the cosmic ray composition changes through the
knee region of the spectrum, becoming heavier with increasing
energy \cite{glasetal99,aretal99,foetal99}. For example, the
CASA-MIA experiment measures the mass composition in the energy
range $10^{14} - 10^{16}$ eV \cite{glasetal99}, and shows that
above $10^{15}$ eV the iron group nuclei should dominate. The
analysis of hybrid data from the High Resolution Fly's Eye
prototype and the MIA muon array indicates that the mass
composition becomes lighter again in the region $10^{17} -10^{18}$
eV \cite{abetal99}, and is proton dominated at around $10^{19}$
eV \cite{bietal93}.

It is usually argued that cosmic rays with energies below
$10^{14}$ eV are accelerated at shocks of expanding supernovae
(see e.g. ref.~\cite{bv00}). It seems that the acceleration
mechanism of the higher energy cosmic rays may be different,
although supernova shocks of various types have also been
proposed \cite{bie93} and also lead to fine structure in the
spectrum and composition in the knee region \cite{SBG93}. This
idea is consistent with the recently observed variation in the
composition of cosmic rays in the region $10^{15} - 10^{16}$eV
mentioned above. However, it is uncertain what fraction of the
cosmic rays above $10^{14}$ eV, are actually accelerated by
supernova shocks, and so other mechanisms able to accelerate
cosmic rays, particularly heavy nuclei, to energies above
$10^{15}$eV may be required. It has also been suggested that
protons and heavy nuclei could be accelerated to such energies in
pulsar winds \cite{osgu69,kow74,ga94,beo99}.

Nuclei can be also accelerated in pulsar magnetospheres as
discussed by Cheng, Ho \& Ruderman~\cite{chr86}, Cheng \&
Chi~\cite{cc96} and Bednarek \& Protheroe~\cite{bp97}.  In our
previous work \cite{bp97,pbl98}, we considered acceleration of
iron nuclei in the magnetospheres of very young pulsars, and
Crab-type pulsars, in the context of $\gamma$-ray and neutrino
emission by supernova remnants.  We found that, although the
photo-disintegration process of iron nuclei during their
acceleration and propagation through the neutron star
magnetosphere is important, heavy nuclei would be injected into
the medium of expanding supernova remnant. In the present paper
we compute the possible contribution of nuclei injected by the
galactic population of gamma-ray pulsars to the observed cosmic
ray spectrum. We show that gamma-ray pulsars can contribute to
the cosmic ray spectrum with energies above $\sim 10^{15}$ eV.

\section{The pulsar spin evolution}

Neutron stars are likely to be born with initial periods
$P_{0}\le 10$ ms if we assume that the angular momentum of the
pre-supernova star is conserved during the collapse.  However, the
initial angular momentum may be lost in the case of hot rapidly
rotating neutron stars by emission of gravitational radiation due
to the r-mode instability \cite{an98,aks98,lom98,oetal98}.
As a result, the star would be left with a period of about 10 ms
at about 1 year after its formation. After that, the period of
the neutron star would change, mainly as a result of emission of
magnetic dipole radiation with power given by
\begin{eqnarray}
L_{\rm em} = B^2 R_{\rm NS}^6 \Omega^4 \sin^2 i/6c^3,
\label{eq1}
\end{eqnarray}

\noindent
where $B$ is the surface magnetic field at the magnetic poles,
$R_{\rm NS}$ is the radius of the neutron star, $i$ is the angle
between rotational and magnetic axes, and $\Omega = 2\pi/P$,
where $P$ is the period of the star.  Also, gravitational
radiation may be important for the case of stars having
significant ellipticity $k$,
\begin{eqnarray}
L_{\rm gr} = 32 G I^2 k^2 \Omega^6/5c^5,
\label{eq2}
\end{eqnarray}
\noindent
where $G$ is the gravitational constant, $I$ is the moment of
inertia \cite{st83}.

We have investigated the population of young pulsars observed
with age $< 10^6$ years (about 100 pulsars) from the catalogue of
Taylor, Manchester \& Lyne~\cite{tml93} in order to estimate the
neutron star ellipticity. By starting with an assumed initial
period $P_0=10$~ms, perpendicular component of magnetic field
$B\sin i$, and ellipticity $k$ we may use Eqs.~\ref{eq1}
and~\ref{eq2} giving the rate of loss of rotational kinetic by
magnetic dipole radiation and gravitational radiation,
respectively, to calculate the period $P$ and period derivative
$\dot{P}$ as a function of time since birth.  For each of the
young pulsars considered, and assuming standard values of $I$ and
$R_{\rm NS}$, and the inferred value of $B\sin i$, we obtain the
value of $k$ such that $P$ and $\dot{P}$ match those
observed~\cite{bs00}.  We note that the values of $k$ obtained
are rather insensitive to the assumed initial period for pulsars
older than $\sim 10^3$ years.  The results are plotted in
Fig.~1, and show that a pulsar's ellipticity must be correlated
with the perpendicular component of its surface magnetic field
multiplied by its radius and divided by its moment of inertia.
Since the theoretical radii and masses of neutron stars lie in a
rather narrow range, we conclude that there exists a correlation
between the ellipticity and the surface magnetic field for
neutron stars. The correlation coefficient obtained is equal to
$\sim 0.65$, which for 100 pulsars means that the correlation is
highly significant. The inferred dependence
of pulsar ellipticity on surface magnetic field can be well
approximated by
\begin{eqnarray}
{\rm log}~k = -22.3 + 1.57 {\rm log}~B\sin i,
\label{eq3}
\end{eqnarray}
\noindent
where $B$ is in Gauss.  

The above result suggests that the ellipticity of a pulsar may be
due to the deformation of the shape of the neutron star by its
strong interior magnetic field (see e.g. \cite{kok99}). The
prescription for pulsar spin evolution we apply in our present
work is based on Eqs.~\ref{eq1}--\ref{eq3}.  Note, however, that
in the case of some specific pulsars, the difference between the
result from Eq.~3 and estimates based on the known age of the
pulsar is significant, e.g., in the case of the Crab pulsar for
which the estimated ellipticity is $\sim 3\times 10^{-4}$
\cite{st83}.  In the case of the recently-discovered pulsar in
the historical (AD 386) SNR G11.2-03, the standard
formula~\cite{tml93} gives for the pulsar $B = 1.7 \times
10^{12}$ G, and using the historical age $t = 1612$ yr a pulsar
initial period of $~62$ ms is found \cite{Torii99}.  If
gravitational radiation were included, the initial period could
be much shorter.  We find that for the same magnetic field, an
ellipticity $k = 1.49\times 10^{-3}$ would give an initial period
of 10~ms for AD 386.  This value of $k$, together with $B = 1.7
\times 10^{12}$ G, is consistent with the correlation described
by Eq.~3 and seen in Fig.~1.

\section{Acceleration of nuclei by young pulsars}

Periodic emission of photons from radio to $\gamma$-ray energies
gives direct evidence of the existence of electric potentials in
neutron star magnetospheres which are able to accelerate
particles to energies of at least $\sim 100$ GeV. To explain
these observations, a number of models have been proposed
including the polar gap model, the slot gap model, and the outer
gap model (see e.g. Arons~\cite{aro96}). In the present paper, we
concentrate on the outer gap model which is more attractive from
the point of view of acceleration of nuclei by pulsars because it
postulates induction of larger electric potentials in the
pulsar's magnetosphere.  The inner gap (polar cap) models
(e.g. Ruderman \& Sutherland~\cite{rs75}, Harding \&
Muslinov~\cite{hm98}) predict that the electric potentials
induced in the inner gaps are significantly lower (of the order
of $\sim 10^{13}$ V), than expected in the outer gaps for pulsars
with the same parameters.  Nuclei accelerated through such
potentials would not suffer efficient photo-disintegration in
collision with thermal photons with a characteristic temperature
$\sim 10^6$ K emitted by the polar cap or from the whole surface
of a classical radio pulsar (only very young pulsars with ages of
days to a month are sufficiently hot to photo-disintegrate Fe
nuclei~\cite{pbl98}).  Hence, we neglect this stage of
pre-acceleration of Fe nuclei during their passage through the
inner gap. Moreover, the inner and outer gaps do not need to
extend along these same magnetic field lines, and so they do not
need to be physically connected.

Nuclei, most probably iron nuclei, can be stripped off the hot
surface of a neutron star due to the strong electric fields and
the bombardment by particles produced in the pulsar magnetosphere
\cite{rs75,as79}. This can occur where the magnetic and rotation
axes are counter-aligned, and therefore where the co-rotation
charge is positive above the poles, and can pull Fe nuclei from
the surface.  If one assumes that pulsars have random
orientations of these axes, then Fe nuclei can be extracted from
a significant fraction of pulsars, but if this is not the case
then our results could be significantly affected.

These nuclei are then accelerated by the pulsar outer gap
electric field to high energies.  We assume that the process of
extraction of iron nuclei from the pulsar surface occurs during
the $\gamma$-ray emitting stage of the pulsar as the production
of $\gamma$-rays in the outer gap requires efficient production
of relativistic $e^\pm$ plasma, and this also heats the polar cap
region. Chen \& Ruderman ~\cite{cr93} distinguish two types of
gamma-ray pulsars, so called Crab-like (Crab-type) and Vela-like
(Vela-type). Beyond the Vela-like death line the gap cannot
close, and this prevents copious production of secondary $e^\pm$
pairs, and causes efficient production of gamma-ray photons to
cease.  However, electric potentials inside the outer gap remain,
and the acceleration of primary particles is still possible.  In
more recent papers (Romani~\cite{ro96}; Zhang \& Cheng
~\cite{zc97}; Cheng \& Zhang~\cite{cz98}) a third type of pulsar,
so called Geminga-like pulsars, is distinguished. In these
pulsars the production of secondary $e^\pm$ pairs inside the
outer gap is sustained by the soft X-ray emission from the polar
cap region.  The death line for Geminga-like pulsars occurs when
there is insufficient flux of thermal photons from the polar cap
region of the pulsar.  However, primary particles can be still
accelerated inside the outer gap even after a pulsar has crossed
the Geminga-like death line (only pair production stops). We
expect that the injection of the Fe nuclei from the region of the
polar cap is related to the thermal radiation field produced by
this region of the neutron star surface.  The thermal radiation
field drops significantly at a neutron star age $\sim 10^6$ yr
(see Fig. 2 in Romani~\cite{ro96}). We estimate that at this age
the pulsar stops injecting Fe nuclei into the magnetosphere.

\subsection{Model for the electric field in the outer gap}
 
During the $\gamma$-ray production stage of pulsars, three
different types of outer gap can be distinguished, depending on
the relative importance of different processes of $\gamma$-ray
production: the Crab-type and Vela-type gaps (Cheng, Ho,
Ruderman~\cite{chr86}), and the Geminga-type gap
(Romani~\cite{ro96}, Zhang \& Cheng~\cite{zc97}). The separation
between the Crab-type and the Vela-type outer gaps depends on the
pulsar period and surface magnetic field, and is given by Eq.~25
of Chen \& Ruderman~\cite{cr93}
\begin{eqnarray}
P_{\rm CV}\approx 0.04 B_{12}^{2/5},
\label{eq5}
\end{eqnarray}
\noindent
where $B_{12}$ is the pulsar surface magnetic field in $10^{12}$ G,
and the period $P_{\rm CV}$ is in seconds. The maximum potential
difference across different types of outer gap can be approximated
by the simple formula (Cheng~\cite{c00}, private communication),
\begin{eqnarray}
\Phi(B,P)\approx 6.6\times 10^{12} f^2 B_{12} P^{-2}~{\rm V}.  
\label{eq6}
\end{eqnarray}
\noindent
This formula gives the potential drop in outer gaps a factor of
two lower than derived in original paper by Cheng, Ho \&
Ruderman~\cite{chr86}, and is more consistent with the recent
works on the outer gap model~\cite{hs99,crz00}.  However, we note
that this simple formula may not match all observational data
well because of uncertainties of viewing angle and inclination
angle which are very important in determining the size of the
outer gap.  The parameter $f$ is used to parametrize the gap
voltage and power, and is defined by Cheng, Ho \&
Ruderman~\cite{chr86}

\begin{eqnarray}
f\approx 82 B_{12}^{-13/20}P^{33/20},
\label{eq7}
\end{eqnarray}
\noindent
Formulae (5) and (6) give the potential difference $\sim 3\times
10^{14}$ eV for the Crab pulsar.
For Geminga-type gaps (Zhang \& Cheng~\cite{zc97})
\begin{eqnarray}
f\approx 5.5 B_{12}^{-4/7}P^{26/21}.
\label{eq8}
\end{eqnarray}
 
For Vela-type outer gaps a simple definition of the parameter $f$
is not available. We assume that in this case the potential drop
in the outer gap is equal to a constant fraction of the total
potential drop in the outer gap given by Eq.~(5) with $f^2=0.15$
for all periods (see \cite{rc88}). The dependence of the
potential difference across the outer gap as a function of the
pulsar period is plotted in Fig.~\ref{fig2} for three different
values of the surface magnetic field of the pulsar.

According to Cheng, Ho \& Ruderman~\cite{chr86} the potential
drop across the gap depends also on the position of injection
into the outer gap field, $z$ ($z$ ranges between 0 and $a$), in
relation to the maximum gap width $a$. Therefore, we assume that
the potential drop for position $z$ is given by (see definitions
below eq.~3.10 in ref.~\cite{chr86})
\begin{eqnarray}
\Phi (z) = 4\Phi(B, P) {{z}\over{a}} [1-{{z}\over{a}}].
\label{eq9}
\end{eqnarray}

\subsection{Radiation field inside the outer gap}

Iron nuclei accelerated in the outer gap potential can suffer
significant photo-disintegration in collisions with photons
produced inside the gap. We estimate the density of soft photons
inside outer gaps of pulsars with a given period and magnetic
field by scaling from the density of soft photons inside the
outer gap of the Crab pulsar assuming the outer gap model. This
density can be estimated from the observed pulsed flux of soft
photons, assuming that the dimension of the outer gap is of the
order of the light cylinder radius. Ho~\cite{h89} has shown, by
analyzing a sample of Crab-type pulsars with periods $\sim15$~ms,
that the luminosity of the photons produced in the gap does not
change drastically for Crab-type outer gaps.  Based on the
results of Ho~\cite{h89}, we interpolate/extrapolate the density
of photons in the present Crab outer gap for other periods.  
The $\gamma$-ray luminosity of pulsars scales as $\sim
f^3$ times the total spin down pulsar luminosity \cite{chr86},
and so the $\gamma$-ray luminosity of the Crab type pulsars
depends on the magnetic field as $\sim B^{1/20}$ (see
Eq.~\ref{eq7}), i.e. it is only very weakly dependent on
$B$. Since energies of synchrotron photons are proportional to
the magnetic field in the gap, then the photon density in
Crab-type outer gaps should be inversely proportional to the
pulsars' magnetic fields.

We note that the observations seem to suggest that the
$\gamma$-ray luminosities of different types of pulsar are
proportional to $B P^{-2}$ (e.g. Fig.~7 in Arons~\cite{aro96}),
although different models predict different dependencies of
$\gamma$-ray luminosities on $B$ and $P$~\cite{h00}. However the
above dependence should be considered with some caution because
it interpolates between the Geminga pulsar (and PSR 1055-52) and
the Vela pulsar (and its twin PSR 1706-44).  The pulsar PSR
1951+32 deviates by an order of magnitude and the Crab pulsar by
a factor of two, and if the distance to the Vela pulsar is $\sim
250$ pc but not $\sim 500$ pc~\cite{csd99}, then Vela pulsar
deviates by a factor of four from this dependence.  Also, the
distance to Geminga pulsar is not well known.

The density of soft photons in Vela-type outer gaps is obtained
based on the observed change of the pulsed luminosity of the Vela
type pulsars (e.g. \cite{rc88}), and our estimate of $f$ in the
case of Vela-type pulsars, with the normalization to photon
density in the Crab-type outer gap at the transition period given
by Eq.~(\ref{eq5}). A good fit to the luminosity of the Vela-type
pulsars is obtained if the density of photons in the outer gap
drops with period as $\propto P^{-3}$. Therefore, we apply the
following model for the soft photon density in Crab and Vela-type
pulsars for different periods and surface magnetic fields
\begin{eqnarray}
n_{\rm ph}\approx n_{\rm ph}^{\rm Crab}\cases 
{ {{P_{\rm Crab}B_{\rm Crab}}\over{P B}}, 
& the Crab-type; \cr
{{P_{\rm Crab}B_{\rm Crab}}\over{P B}} 
\left({{P_{\rm CV}}\over{P}}\right)^3, 
& the Vela-type,\cr} 
\label{eq10}
\end{eqnarray}
\noindent
where $P_{\rm Crab} = 0.033$ seconds, $B_{\rm Crab}=4\times 10^{12}$ G
and $n_{\rm ph}^{Crab}$ is the density of photons in the outer gap of 
the Crab pulsar at present \cite{chr86}(Cheng, Ho \& Ruderman~1986).

The Geminga-type outer gap is determined by the thermal photons
produced on the neutron star surface. This thermal radiation
field for pulsars with ages corresponding to Geminga-type pulsars
is too low to create a significant target for
photo-disintegration of Fe nuclei.  The epoch of gamma-ray
emission of Geminga-type pulsars is determined by the temperature
of the neutron star surface which, according to models of neutron
star evolution, drops rapidly after $\sim 10^6$ years (see
e.g. Fig.~2 in Romani~\cite{ro96}). Therefore, Geminga-type outer
gaps should not be able to operate after $\sim 10^6$ years from
the explosion of supernova. Here we assume that the process of
extraction of Fe nuclei from the neutron star surface does not
occur after this age because, as we noted above, it is probably
caused by heating of the polar cap region by $e^\pm$ pairs
created in the pulsar magnetosphere.

\subsection{Number of nuclei accelerated in the outer gap}

As in our previous paper \cite{bp97}, we assume
that the number of Fe nuclei injected 
per unit time, $\dot{N}_{\rm Fe}$, is simply related to the
total power output of the pulsar $L_{\rm em}(B,P)$~\cite{mt77},
\begin{eqnarray}
\dot{N}_{\rm Fe}= \xi L_{\rm em}(B,P)/Ze \Phi(B,P),
\label{eq11}
\end{eqnarray}
\noindent
where $\xi$ is a free parameter describing the fraction of the
total power taken by relativistic nuclei accelerated in the outer
gap, $Z=26$ is the atomic number of Fe, and $\Phi(B,P)$ is
defined by Eq.~\ref{eq6}. The parameter $\xi$ has to be always
less than one since the power of accelerated Fe nuclei can never
be greater than the spin-down power of the pulsar.  The charge
density estimated from Eq.~\ref{eq11} exceeds the `classical'
Goldreich-Julian charge density close to the polar caps $\rho =
-(\Omega B / 2\pi c) [1 - (\Omega r/c)^2 \sin^2\theta]^{-1}
\approx - \Omega B/2\pi c,$ where the approximate formula is
valid close to the surface where $\Omega r \sin\theta \ll
c$. However, as noted by Goldreich \& Julian~\cite{gj69} below
their Eq. (8), the above formula is only valid in the co-rotation
portion of the magnetosphere. If there is additional negative
current onto the surface accompanying the charge injected from
the surface in the form of Fe nuclei, then the above formula is
not valid.  Positive charge may flow onto the surface along
the same magnetic lines as Fe nuclei, and could be produced
somewhere in the pulsar magnetosphere, not necessary close to the
pulsar's surface (e.g. possibly in the outer gap).  Pair
production just above the polar cap does not need to occur in
this case.

\subsection{Number of neutron stars with different
parameters in the Galaxy}

As discussed above, we assume that all pulsars are born with
periods of about 10 ms, or that they reach this period soon after
formation by emission of gravitational waves due to the r-mode
instability.  As we argued in Section~2, pulsars need to have a
significant ellipticity $k$ in order to reach their observed
periods from a period at birth of 10 ms.

The distribution of the pulsar magnetic fields can be evaluated
from observations, assuming that the surface magnetic field does
not decay significantly with pulsar age. For example,
Narayan~\cite{nar87} approximates the distribution of surface
magnetic fields of pulsars at birth by
\begin{eqnarray}
dN/d(\log B)\approx 0.065 (10^{12} G/ B)~{\rm yr}^{-1}, 
\label{eq14}
\end{eqnarray}
\noindent
for $B > 2\times 10^{12}$ G. We estimate the distribution of
pulsars with surface magnetic fields below $2\times 10^{12}$ G
based on the Fig.~(13) in Narayan~\cite{nar87}. It is likely that
this approximation underestimates the number of pulsars with
relatively low magnetic fields because of observational selection
effects. The above formula is normalized to a formation rate of
pulsars in the Galaxy equal to 1 per 70 years.

\section{Acceleration and propagation of nuclei from the neutron star 
surface}

Nuclei extracted from the neutron star surface are accelerated
and photo-disintegrated in the outer gap. Their disintegration
products, and those which escape without any disintegration, are
injected into the nebula and lose energy in collisions with
matter ejected during the supernova explosion, and as a result of
the expansion of nebula. All these processes are discussed below.

\subsection{Propagation of nuclei through pulsar magnetospheres}

The importance of photo-disintegration of nuclei with different
mass numbers during propagation of nuclei in the radiation field
of the outer gap is determined by the reciprocal mean free path,
which we calculate using cross sections of Karaku\l a \&
Tkaczyk~\cite{kt93} (with improvements given in
ref.~\cite{pbl98}). The results of such a calculation for an
outer gap with parameters applicable to the Crab pulsar are shown
in Fig.~1 of Bednarek \& Protheroe~\cite{bp97}.  It is evident
that multiple photo-disintegrations of primary Fe nuclei will
occur for this pulsar. In our present calculations, the
photo-disintegration of nuclei in the thermal radiation field
emitted by the pulsar surface is neglected since it may be only
important during a few days after creation of the pulsar when its
surface temperature is of the order of $10^7$ K \cite{pbl98}, and
consider only photodisintegration in the outer gap's non-thermal
radiation field.

We simulate the photo-disintegration process of nuclei
accelerated by pulsars using the evolution model, initial
parameters of pulsars, the outer gap model for acceleration, and
the radiation field discussed in Sections 2 and 3. Multiple
photo-disintegrations occur only for pulsars with Crab and
Vela-type outer gaps. We simulate the acceleration and
propagation of iron nuclei in pulsar outer gaps in order to
obtain the energy spectrum of neutrons and protons (extracted
from Fe nuclei) and the spectra of escaping secondary
nuclei. Note, however, that the fate of neutral particles
(neutrons) and charged particles (protons and secondary nuclei)
is different. The neutrons move ballistically through the
surroundings of the pulsar, and decay at distances determined by
their Lorentz factors, and this can be inside or outside of the
expanding envelope of the supernova. We compute the spectra of
neutrons (and so protons from their decay) which decay inside and
outside the Nebula as in Bednarek \& Protheroe \cite{bp97}. In
the case of protons from neutrons decaying outside the nebula, we
assume that all of them escape into the Galactic medium without
being overtaken by the expanding nebula (we drop the second term
in brackets in Eq.~(5) in Bednarek \& Protheroe~\cite{bp97} which
represents the effects of protons from neutrons decaying outside
the nebula being overtaken and trapped by the expanding nebula at
some later time). This term can be neglected since the minimum
diffusion distance of these protons, with Lorentz factor
$\gamma_{\rm p} = 10^5$, in the Galactic magnetic field $5\times
10^{-6}$ G is comparable to the radius of nebula during its time
of free expansion \cite{bp97}.

Protons and secondary nuclei injected into the Nebula are trapped
by the magnetic field of the nebula, and suffer adiabatic and
collisional energy losses during the time of expansion of the
nebula. These effects are considered in the next subsection.

\subsection{Interaction of nuclei with matter inside nebulae}

Soon after the supernova explosion, when the nebula was
relatively small, nearly all the energetic neutrons would be
expected to decay outside the nebula.  However, at early times we
must take account of collisions with dense matter of the
expanding nebula.  The optical depth for neutrons moving
ballistically through the envelope may be estimated from
$\tau_{nH}\approx \sigma_{np} n_H r \approx 8.6\times 10^{14} M_1
v_8^{-2} t^{-2}$, where $M = M_1M_{\odot}$ is the mass ejected
during the supernova explosion in units of solar masses, $r=vt$,
$n_H = M/(4/3\pi r^3 m_p)$ is the number density of target
nuclei, and $v = 10^8 v_8$ cm s$^{-1}$ is the expansion velocity
of the nebula.  We note that $\tau_{nH} = 1$ at $t \approx 0.93
M_1^{1/2}v_8^{-1}$~yr.

Charged nuclei (protons, protons from decay of neutrons, and
secondary nuclei) injected by the pulsar are captured by the
nebula's magnetic field and interact with the dense material of
the expanding supernova remnant. The optical depth for these
nuclei depends on the time, $t$, of the injection of nuclei, and
can be estimated from
\begin{eqnarray}
\tau_{\rm Ap}(t)\approx 1.3\times 10^{17} A^{2/3}M_1v_8^{-3}t^{-2}. 
\label{eq15}
\end{eqnarray}
Interactions of nuclei with matter inside the nebula become
negligible when the optical depth becomes less than one. For the
parameters derived for the Crab, $v_8 = 2$ and $M_1 =3$, and the
average value of the mass number of nuclei injected from the
pulsar magnetospheres into the nebula $<A>\approx 48$, the nebula
becomes transparent for nuclei at times $t > 3.6\times 10^8
A^{1/3}M_1^{1/2} v_8^{-3/2}\approx 22$ years. Nuclei injected at
earlier times should suffer significant fragmentation in
collisions with the matter inside the nebula. We have simulated
this process of fragmentation of nuclei injected by the pulsar
applying the semi-empirical formulae for the cross sections for
the fragmentation of nuclei of Silberberg \& Tsao~\cite{st73}
with parameters given in their tables 1A,B,C,D. In our
simulation, we assumed that in a single collision a nucleus
fragments into two nuclei with different mass numbers.

Protons created during complete fragmentation of nuclei are
considered as being captured by the magnetic field of the
nebula. In contrast, neutrons which move ballistically through
the nebula have some chance of decaying inside or outside the
nebula. These effects have been included in our simulation
code. We consider the interactions of these secondary protons and
neutrons with matter inside the nebula in the same way as for
protons and neutrons from photo-disintegration of iron nuclei in
the pulsar's magnetosphere.

\subsection{Escape of nuclei from nebulae}

We assume that nebulae expand during the free expansion phase
with the velocity observed in the case of the Crab Nebula,
i.e. 2000 km s$^{-1}$. When the amount of matter ejected during
the supernova explosion becomes comparable to the amount of
interstellar matter swept up by the nebula, then the nebula
reaches the Sedov phase.  The Sedov phase starts when the nebula
reaches radius
\begin{eqnarray}
R_{\rm S} = 6.8\times 10^{18} (M_{1}/n)^{1/3}~{\rm cm},
\label{eq16}
\end{eqnarray}

\noindent
and this occurs at time 
\begin{eqnarray}
t_{\rm S} = R_{\rm S}/v,
\label{eq17}
\end{eqnarray}
\noindent
after the explosion.  Here $n$ is the density of the surrounding
medium (cm$^{-3}$), and $v$ is the expansion velocity of
supernova during its phase of free expansion.  For a supernova
with an ejecta mass of $M_{\rm SN} = 3 M_{\odot}$, a velocity of
free expansion $v = 2000$ km s$^{-1}$ as observed for the Crab
Nebula, and a density of the interstellar medium $n = 0.3$
cm$^{-3}$, the Sedov phase should begin when the nebula reaches a
radius of $R_{\rm S}\approx 4.8$ pc at about $t_{\rm S}\approx
2000$ years. This is only about twice as large as the present
radius (and age) of the Crab Nebula.  We assume the following
approximation for the radius of a nebula at time $t$

\begin{eqnarray}
R_{\rm Neb}(t) = \cases { v t, &if $t \le t_{\rm S}$; \cr
 v t_{\rm S}^{3/5} t^{2/5}, &if $t > t_{\rm S}$. \cr}
\label{eq18}
\end{eqnarray}

The magnetic field inside the nebula changes with time and
depends on the distance from the center of the nebula, i.e. the
pulsar.  We estimate it applying the results of the Kennel \&
Coroniti~\cite{kc84} model, which gives the distribution of the
magnetic field inside the nebula as a function of the parameter
$\sigma$ (the ratio of the magnetic energy flux to the particle
energy flux at the location of the pulsar wind shock at radius
$R_{\rm sh}$). We estimate the magnetic field inside the nebula
(in units of the magnetic field strength $B_{\rm sh}$ at the
shock location) as a function of its radius based on the figs. 3
and 7 in Kennel \& Coroniti~\cite{kc84}.  For nebulae with
$\sigma \gg 0.003$, the magnetic field drops significantly with
the distance from the center of the nebula (fig.~3 in
~\cite{kc84}), and can reach a value typical for the interstellar
medium ($5\times 10^{-6}$ G) at distance $R_{\rm B}$ which is
less than the radius of the expanding nebula $R_{\rm Neb}$, and
so we assume that beyond radius $R_{\rm B}$, the magnetic field
is equal to $5\times 10^{-6}$ G.

The magnetic field at the shock, $B_{\rm sh}$, can be found from
the dependence of the pulsar's magnetic field in the region of
the inner magnetosphere ($B\propto R^{-3}$), and the pulsar wind
zone ($B\propto R^{-1}$). It is given by

\begin{eqnarray}
B_{\rm sh} \approx \sqrt{\sigma} B_{\rm s} 
\left({{R_{\rm s}}\over{R_{\rm L}}}\right)^3 
\left({{R_{\rm L}}\over{R_{\rm sh}}}\right), 
\label{eq18a}
\end{eqnarray}

\noindent
where $R_{\rm s}$ and $B_{\rm s}$ are the radius and the surface
magnetic field of the pulsar, and $R_{\rm L} = cP/2\pi$ is the
light cylinder radius of the pulsar. The value of $\sigma$ is
estimated for the Crab pulsar to be $\sim 0.003$ (e.g. Kennel \&
Coroniti~\cite{kc84}). However, it has recently been found that
for the Vela pulsar the value of $\sigma$ is rather close to 1
(Helfand et al.~\cite{hgh00}). To our knowledge, there is no
available information on $\sigma$ for other plerions. Therefore,
in order to fulfill observational constraints of these two
pulsars, we assume that pulsars with periods lower than the
period of the Crab pulsar have $\sigma\approx 0.003$ and pulsars
older than the Vela pulsar have $\sigma\approx 1$.  For pulsars
having a rate of the rotational energy loss between that of the
Crab and Vela pulsars, $\sigma$ is obtained by linear
interpolation between these two values.

The location of the wind shock, $R_{\rm sh}$, can be estimated
from the comparison of the wind energy flux at the shock with the
pressure of the outer nebula, which is determined by the supply
of the magnetic energy to the nebula by the pulsar over its all
lifetime (Rees \& Gunn~\cite{rg74}). The location of the shock at
an arbitrary time $t_{\rm p}$ after explosion of the supernova
can be estimated from

\begin{eqnarray}
{{L_{\rm em}(t_{\rm p})}\over{4\pi R_{\rm sh}^2c}}
\approx {{\int_{0}^{t_{\rm p}} \sigma L_{\rm em}(t) dt}\over{
{{4}\over{3}}\pi R_{\rm Neb}^3}}. 
\label{eq18b}
\end{eqnarray}

As an example, we show in Fig.~3 the characteristic radii in the
nebula, $R_{\rm Neb}$, $R_{\rm sh}$, and $R_{\rm B}$, for the
following parameters: $B_{\rm s} = 3\times 10^{12}$ G, $k =
3\times 10^{-4}$, $v = 2000$ km s$^{-1}$, $M_{\rm SN} = 3
M_{\odot}$, and $n = 0.3$ cm$^{-3}$.  The location of the shock,
estimated from the above formula, is consistent with the
observations of the shock in the Crab~\cite{kc84} and
Vela~\cite{hgh00} nebulae. For older nebulae $R_{\rm B}$ becomes
less than $R_{\rm Neb}$, i.e. the magnetic field inside the outer
regions of expanding nebula drops to $5\times 10^{-6}$ G.

Charged particles, nuclei and protons, injected into the nebula
will diffuse, and finally escape from it into interstellar space.
We assume that particles escape from a nebula at time $t_{\rm
max}$ when the diffusion distance of particles injected at the
time $t$ becomes comparable to the radius of the nebula $R_{\rm
Neb}$ at the time $t_{\rm max}$. In order to estimate this time
at which they escape from the nebula, we compare the diffusion
distance for nuclei with the mass number $A$ and the Lorentz
factor $\gamma_{\rm A}$ in the time dependent magnetic field
inside an expanding nebula with radius given by
Eq.~\ref{eq18}. The distance diffused depends on the diffusion
coefficient which we take to be
\begin{eqnarray}
D=D_0 \rho^{1/3},
\label{eq20}
\end{eqnarray}
as appropriate to a Kolmogorov spectrum of turbulence.  Here,
$\rho\approx E/eZ$ is the particle rigidity, and $D_0$ is
normalized in such a way that $D = r_{\rm L}c/3$ when the Larmor
radius, $r_{\rm L}$, is equal to the radius of the remnant
(largest scale) at time t (given by Eq.~\ref{eq18}).  This is a
conservative assumption as it will give rise to the greatest
trapping of nuclei in the nebula (and highest adiabatic losses)
while being consistent with a Kolmogorov spectrum of turbulence.
However, the precise form adopted for $D$ has little effect on
our results as most particles are released at $t_{\rm max}$. The
diffusion distance is obtained by integration over time from the
injection of nuclei at $t_{\rm inj}$ up to escape at $t_{\rm
esc}$ %
\begin{eqnarray}
r_{\rm dif} = \int_{t_{\rm inj}}^{t_{\rm esc}}
\sqrt{{{3D}\over{2t'}}}  dt'.  
\label{eq21}
\end{eqnarray}
\noindent
Note that the diffusion coefficient depends on the age of the
nebula, and this is included when integrating Eq.~\ref{eq21}.
For nuclei with sufficiently low energies, $t_{\rm max}$ may be
shorter than $t_{\rm esc}$, and we then take $t_{\rm esc} =
t_{\rm max}$.  The adiabatic losses of particles inside nebula
are considered up to the moment when the expansion velocity of
the nebula during the Sedov phase, $v_{\rm exp} = v (t_{\rm
S}/t)^{3/5}$, drops to a few tens km $s^{-1}$ (i.e. to the
characteristic velocities of the clouds in the interstellar
medium). This happens at the age of $\sim 10^6$ yr.  Note that at
this age the magnetic field inside the nebula is larger than
$5\times 10^{-6}$ G only at small region above the radius of the
shock ($R_{\rm B}\approx 2 R_{\rm sh}$, see Fig.~3).

Adiabatic and interaction energy losses of nuclei during the
expansion of the nebula are included by estimating the Lorentz
factor of nuclei at arbitrary time, $t$,
\begin{eqnarray}
\gamma (t) = \gamma (t_{\rm inj})
{{t_{\rm inj} + t}\over{2t_{\rm inj}K^{\tau_{\rm Ap}(t)}}},
\label{eq22}
\end{eqnarray}
\noindent
where $K$ is the inelasticity coefficient, assumed equal to 0.5
for protons and neutrons. For nuclei with mass number $A$, we
take the inelasticity coefficient to be $K\approx 0.5/A^{1/3}$
since the probability of a collision of a single nucleon within
the energetic nucleus, with a proton of the remnant nebula is, on
average, proportional to $(\sigma_{Ap}/A)\propto (A^{2/3}/A)\propto
A^{-1/3}$.

\section{Injection rate of nuclei into the Galaxy}

Taking into account all of the effects discussed above, we
compute the injection rate of nuclei of different mass numbers by
the galactic population of pulsars into the interstellar medium,
and this is given in Fig.~\ref{fig4}(a).  The spectrum shows two
distinct maxima corresponding to heavy nuclei (dot-dot-dot-dash
curve peaking at a few $10^{15}$eV) and to protons (solid curve
peaking at $\sim 10^{14}$eV). The contribution of medium mass
nuclei is visible at energies between the proton maximum and the
heavy nuclei maximum. The main contribution to the spectrum of
nuclei below the high-energy peak is from Crab-type pulsars. This
is clearly seen in Fig.~\ref{fig4}(b) in which we present the
contribution of pulsars having different types of outer gap to
the total spectrum of nuclei injected into the Galaxy the from
pulsars. The main contribution to the maximum at $3 \times
10^6$~GeV comes from Vela-type and Geminga-type pulsars. However
the nuclei with the highest energies are injected by the Vela and
Crab-type pulsars as expected from inspection of the electric
potentials shown in Fig.~\ref{fig2}.

\section{Cosmic ray propagation}

We assume that the radial distribution of the cosmic ray emission
from gamma-ray pulsars follows the distribution of old neutron
stars in the disk of the galaxy as given by Paczynski \cite{pacz90}
\begin{equation}
q(r)2 \pi r \,  dr = a_R \exp(-r/R_{\rm exp}) R_{\rm exp}^{-2} 2 \pi r
\, dr
\end{equation}
with $a_R=1.0683/2\pi$, $R_{\rm exp}=4.5$~kpc, and $r$ being the
the distance from the centre of the Galaxy of radius $r_{\rm
max}=20$~kpc, and that they are uniformly distributed in a disk
of thickness $2 h_g$.  For $h_g$ we take an estimate of the scale
height of pulsar birth locations, $h_g=130$~pc
\cite{CordesChernoff98}, since random pulsar velocities $\sim
200$ km s$^{-1}$ would not significantly increase the scale
height for pulsars much younger than $10^6$ y.  Combining $q(r)$
with the calculated spectra of cosmic rays of atomic number $Z$,
$Q_Z(E)=dN/dE/dt$ presented in Fig.~\ref{fig4}(a), we obtain the
spatial distribution of cosmic ray injection
\begin{equation}
\zeta (r,E) = q(r)Q_Z(E)/(2 h_g).
\end{equation}

We assume that at energies above $10^3$~GeV we can reasonably
neglect interactions and energy losses of cosmic ray nuclei, and
that diffusion is the dominant process in shaping the observed
cosmic ray spectrum.  We consider two different models (Ptuskin
et al \cite{ptuetal99}) for propagation of cosmic rays in the
galaxy which are consistent with the observed secondary to
primary ratios: (i) the minimal reacceleration model, and (ii)
the standard diffusion model.  We use the best-fitting parameters
found by Ptuskin et al \cite{ptuetal99} for a halo half-thickness
$h_h= 5$~kpc (we note here that the cosmic ray intensity at Earth
obtained is rather insensitive to the value of $h_g$
chosen provided that $h_g \ll h_h$).  For the minimal
reacceleration model, the diffusion coefficient used is
\begin{equation}
D =3.8 \times 10^{29} (\rho / {\rm 1 \, GV})^{1/3} \;\;\;\; {\rm
cm^2s^{-1}}
\label{eq:D_kolmogorov}
\end{equation}
where $\rho \approx E/Z$ is the particle rigidity.
For the standard diffusion model, the diffusion coefficient
used is
\begin{equation}
D =3.0 \times 10^{28} (\rho / {\rm 1 \, GV})^{0.54} \;\;\;\; {\rm
cm^2s^{-1}}.
\label{eq:D_standard}
\end{equation}
Assuming cylindrical geometry, we then use Eq. 3.22 of Berezinsky
et al. \cite{betal90}, to calculate the cosmic ray density at
Earth assumed to be in the Galactic plane 8~kpc from the centre
of the Galaxy.

The spectra after propagation are plotted in
Fig.~\ref{fig:propspec0} for the various species where they are
compared with the world data set of observed cosmic ray intensity
\cite{GS98}.  Fig.~\ref{fig:propspec0}(a) and (b) show results
for the two propagation models.  The results are rather
insensitive to the propagation model used, and this can be easily
understood after inspecting the injection spectra in
Fig.~\ref{fig4}(a) and
Eqs.~\ref{eq:D_kolmogorov}--\ref{eq:D_standard}.  The peak at
$10^5$~GeV is due to protons with rigidity $10^5$~GV, and the
peak at $\sim 3 \times 10^6$~GeV is due to iron nuclei with
rigidity $1.5 \times 10^5$~GV, approximately the same as the
proton peak.  Looking at Eqs.~\ref{eq:D_kolmogorov} and
\ref{eq:D_standard}, we note that the diffusion coefficients for
the two propagation models are equal at $1.8 \times 10^5$~GV, and
this is within less than a factor of 2 of the rigidities of both
the proton peak and the iron peak, and so the two propagation
models will give almost identical fluxes for the same pulsar
injection spectrum.  The propagated fluxes are interestingly
close to those observed, with the region between $3 \times
10^6$~GeV and $2 \times 10^7$~GeV being about $20$\% of the
observed flux.

\section{Discussion and Conclusion}

As can be seen in Fig.~\ref{fig:propspec0}, we can expect fine
structure in the spectrum in the region of the knee, as well as
subtle composition changes.  Our calculations were made assuming
100\% efficiency of conversion of spin-down power into
accelerated iron nuclei, i.e.  $\xi=1$ in Eq.~\ref{eq11}.  Other
uncertainties entering our calculation relate the propagation of
cosmic rays.  The diffusion coefficients assumed were based on an
analysis of lower energy cosmic ray data, and their extrapolation
to higher rigidities may be questionable.  This would increase
our predicted flux if the actual diffusion coefficient were
lower.  Also, the pulsar formation rate is rather poorly known.
With all these uncertainties, we feel justified is adopting
$\xi\sim 1$ in order see whether injection of cosmic rays
accelerated in pulsar outer gaps could be potentially
interesting.  

We shall next examine the composition in the knee region of the
spectrum. The mass composition below $\sim 10^{15}$ eV is
characterized by an average value of $<\ln A>\approx 1.2$
\cite{glasetal99}. If we assume that this component of the
spectrum extends to higher energies, and add to it the
contribution of heavy nuclei (with average $A\approx 45$) as
obtained for our pulsar acceleration model, this would change the
average value of $<\ln A>$ to $\sim 2.4$ above the knee.  Such a
change in $<\ln A>$ has recently been reported for the CASA-MIA
air shower array data \cite{glasetal99}.

Our calculations have assumed a steady-state production of
pulsar-accelerated cosmic rays throughout the Galaxy with a
smooth spatial distribution of sources following that inferred
for pulsars.  It neglects the effect of relatively nearby
sources, such as any within 1~kpc releasing cosmic rays during
the past $10^4$~years which would enhance the pulsar
contribution, and so the effects of pulsar acceleration on the
cosmic ray composition could be even more dramatic.  Finally, we
conclude that pulsar-accelerated cosmic rays could make a
significant contribution of heavy nuclei to the cosmic rays
observed in the knee region, and that this would manifest itself
as fine structure in the spectrum, and a heavy composition at $3
\times 10^6$--$2 \times 10^7$~GeV.

\section*{Acknowledgements}

We would like to thank the referee for useful comments and suggestions,
K.S. Cheng for information on the details of the outer gap model, and 
T.A. Porter for discussions on cosmic ray propagation.  The work of WB 
is supported by the {\it Komitet Bada\'n Naukowych} through the grants 
2P03D 001 14 and 2P03C 006 128. The research of RJP is supported by 
the Australian Research Council.

\newpage

\begin{figure}
\epsfig{file=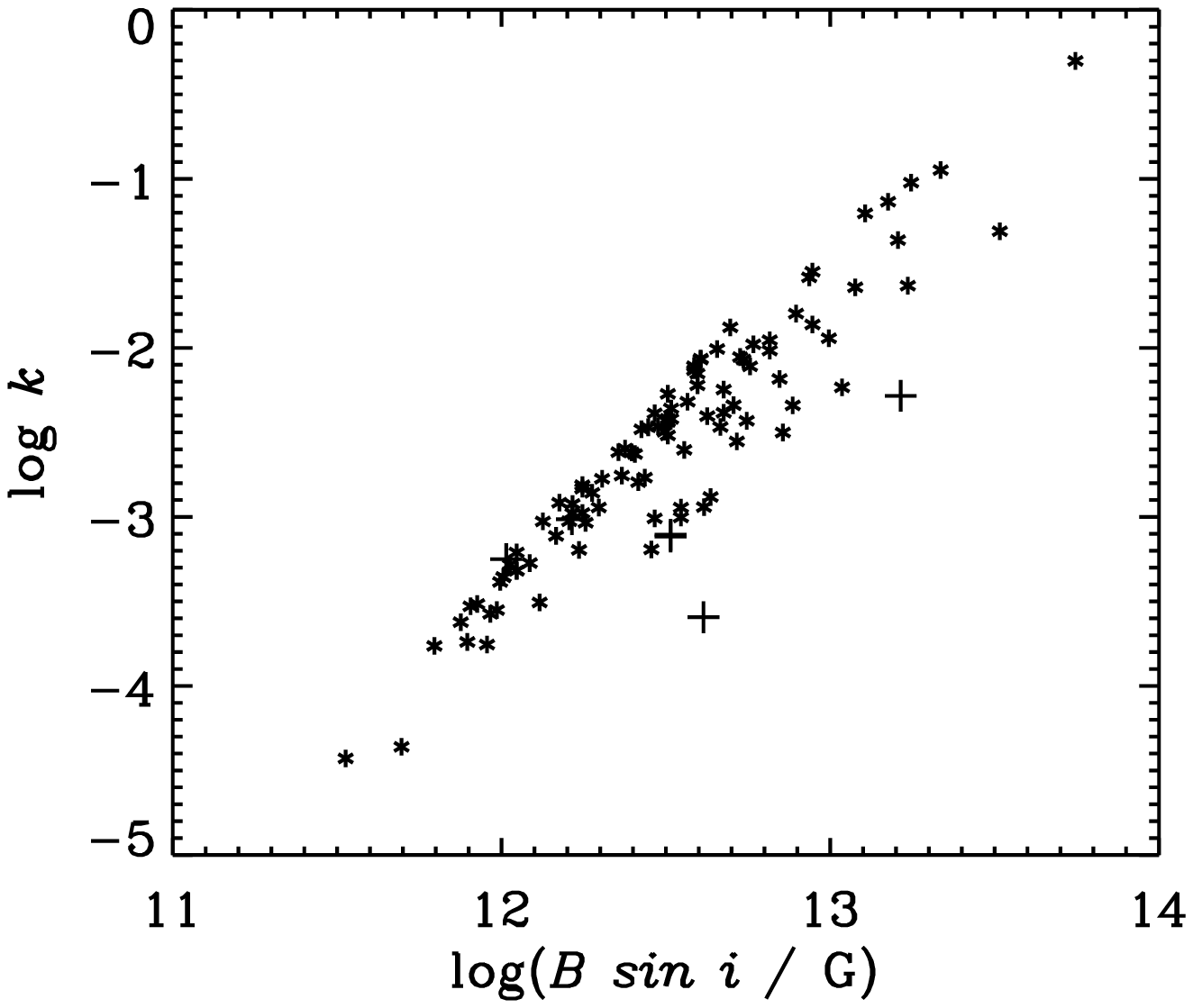, width=18cm}
\caption{Correlation between the minimum values of ellipticity
$k$ of the pulsars with age $< 10^6$ years and their surface
magnetic fields.  The observed $\gamma$-ray pulsars are marked by
crosses and other young pulsars are marked by stars. The initial
periods of all pulsars have been fixed at $P_0 = 10$ ms.}
\label{fig1}
\end{figure}

\begin{figure}
\epsfig{file=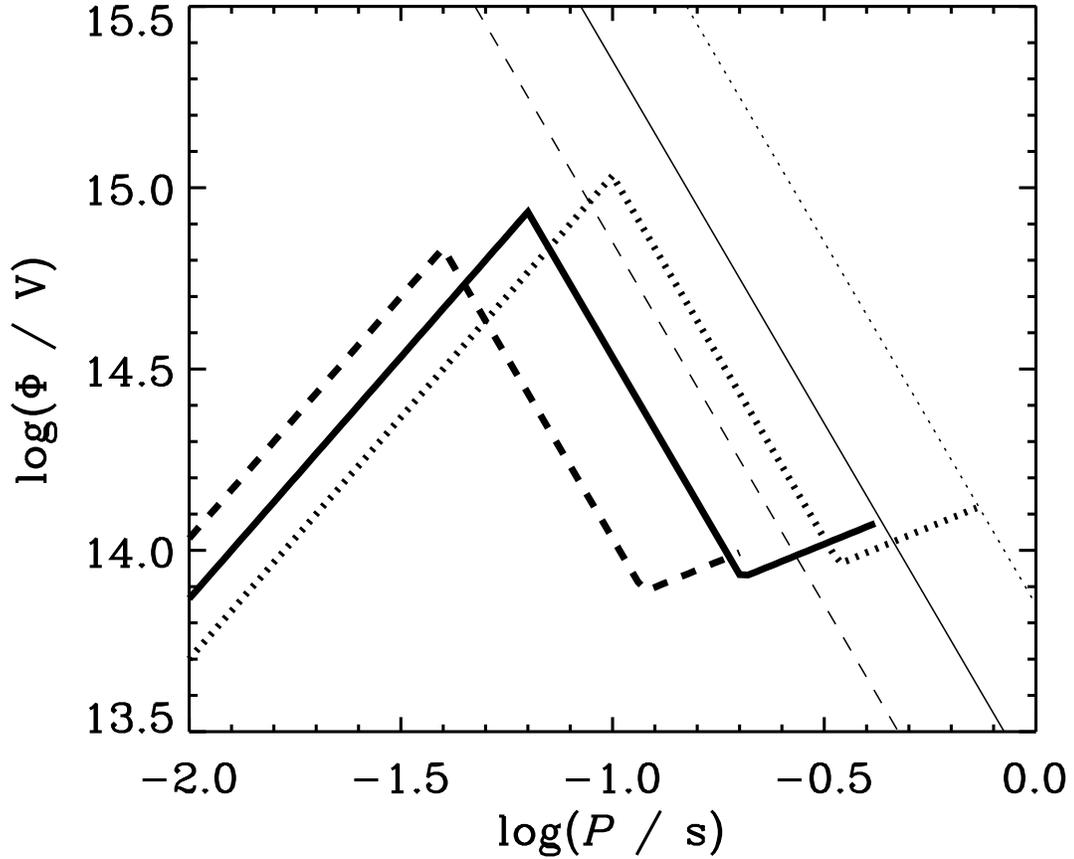, width=18cm}
\caption{Dependence of the potential drop in the outer gap as a
function of pulsar period for three values of surface magnetic
field $B = 10^{12}$ G (dashed curve), $10^{12.5}$ G (full curve),
and $10^{13}$ G (dotted curve). The corresponding thin lines show
the maximum potential drop available across the region of the
polar cap for three values of the magnetic field discussed above.
For a given $B$ value, Crab-type pulsars have the shortest
periods, Vela-type pulsars have intermediate periods, and
Geminga-type pulsars have the longest periods.}
\label{fig2}
\end{figure}

\begin{figure}
\epsfig{file=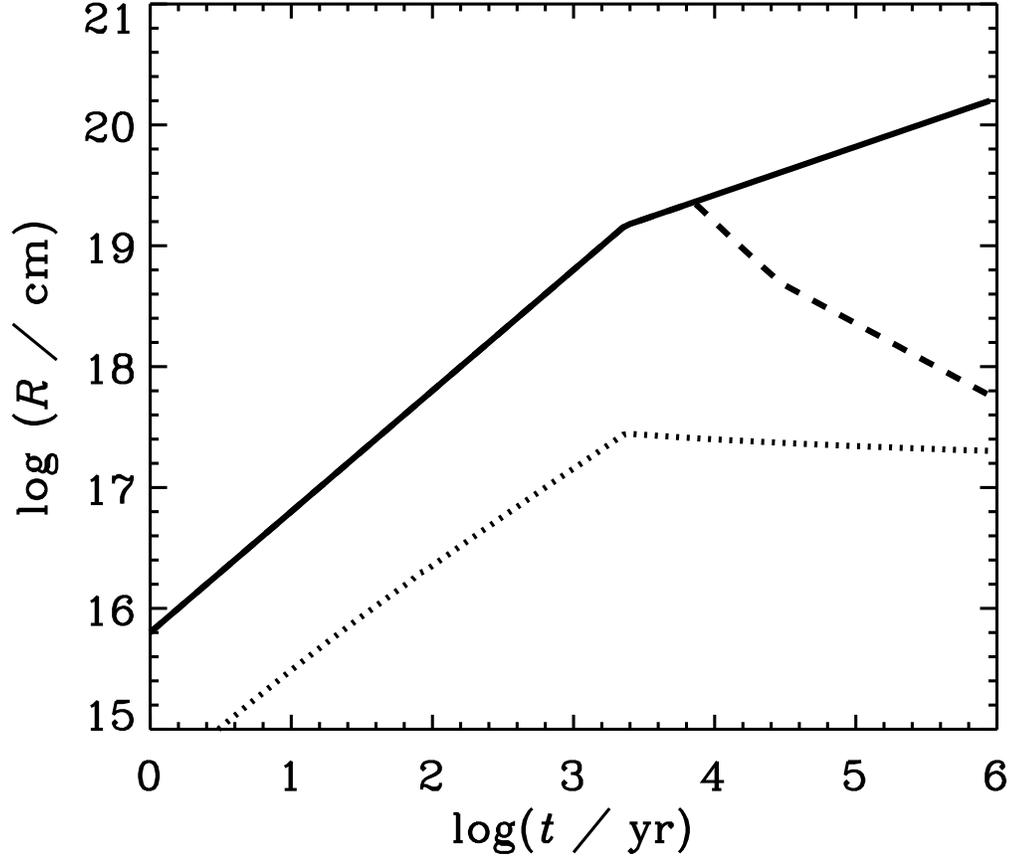, width=18cm}
\caption{Radius of pulsar nebula (full line) as a function of age
for a the pulsar with the parameters expected for the Crab pulsar
and Nebula: surface magnetic field $B = 3\times 10^{12}$,
ellipticity $k\sim 3\times 10^{-4}$~\cite{st83}, initial
expansion velocity $v = 2000$ km s$^{-1}$, mass of the supernova
$M_{\rm SN} = 3 M_{\odot}$, and density of the surounding medium
$n = 0.3$ cm$^{-3}$.  The dashed line shows the radius, $R_{\rm
B}$, above which the magnetic field drops to the value of the
interstellar magnetic field (assumed $5\times 10^{-6}$ G), and
the dotted line shows the radius of the inner shock, $R_{\rm
sh}$, produced by the interaction of the pulsar wind with the
surounding nebula.}  \label{fig3}
\end{figure}

\begin{figure}
\epsfig{file=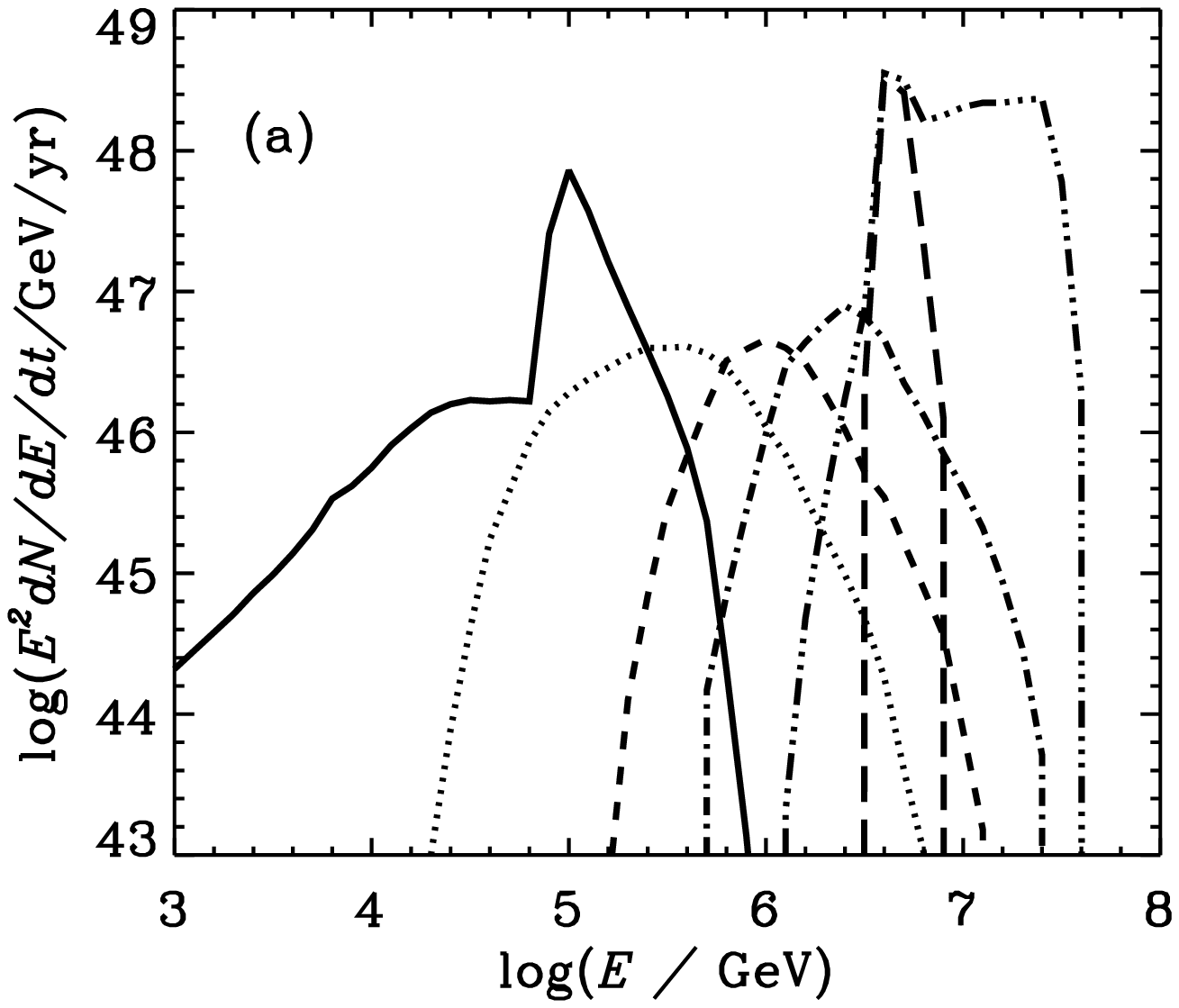, width=9cm}
\epsfig{file=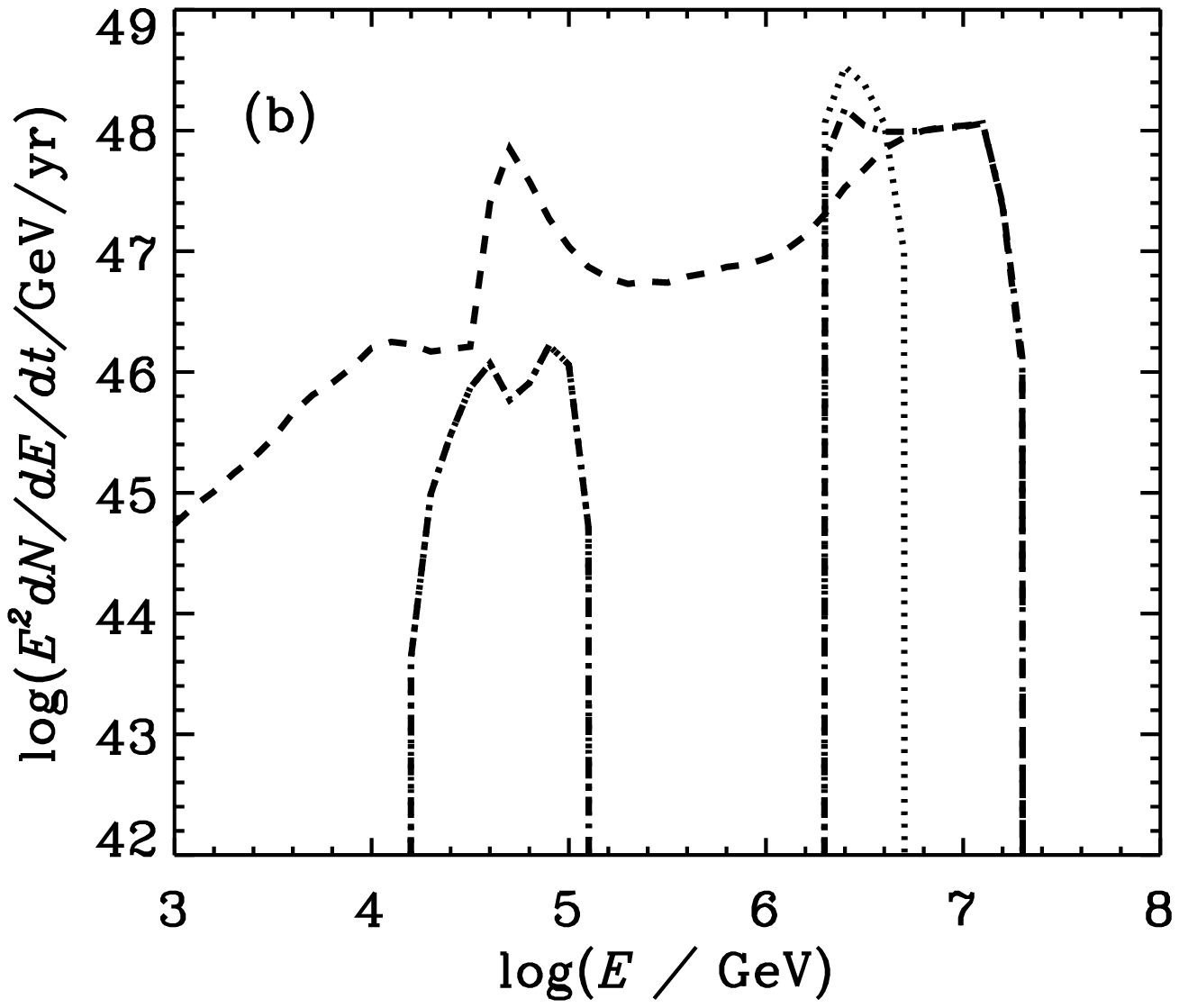, width=9cm}
\caption{Differential spectrum of nuclei injected by pulsars into
the interstellar medium, multiplied by the square of energy per
nucleus.  (a) Contribution of different groups of nuclei: solid
curve -- A=1; dotted curve -- A=2--10; short dashed -- A=11--20;
dot-dashed -- A=21--40; dot-dot-dot-dashed -- A=41--56 (the
contribution of A=56 is also shown separately as the long dashed
curve).  (b) Contribution of Crab (dashed curve), Vela
(dot-dashed curve), and Geminga (dotted curve) type pulsars.}
\label{fig4}
\end{figure}

\begin{figure}
\epsfig{file=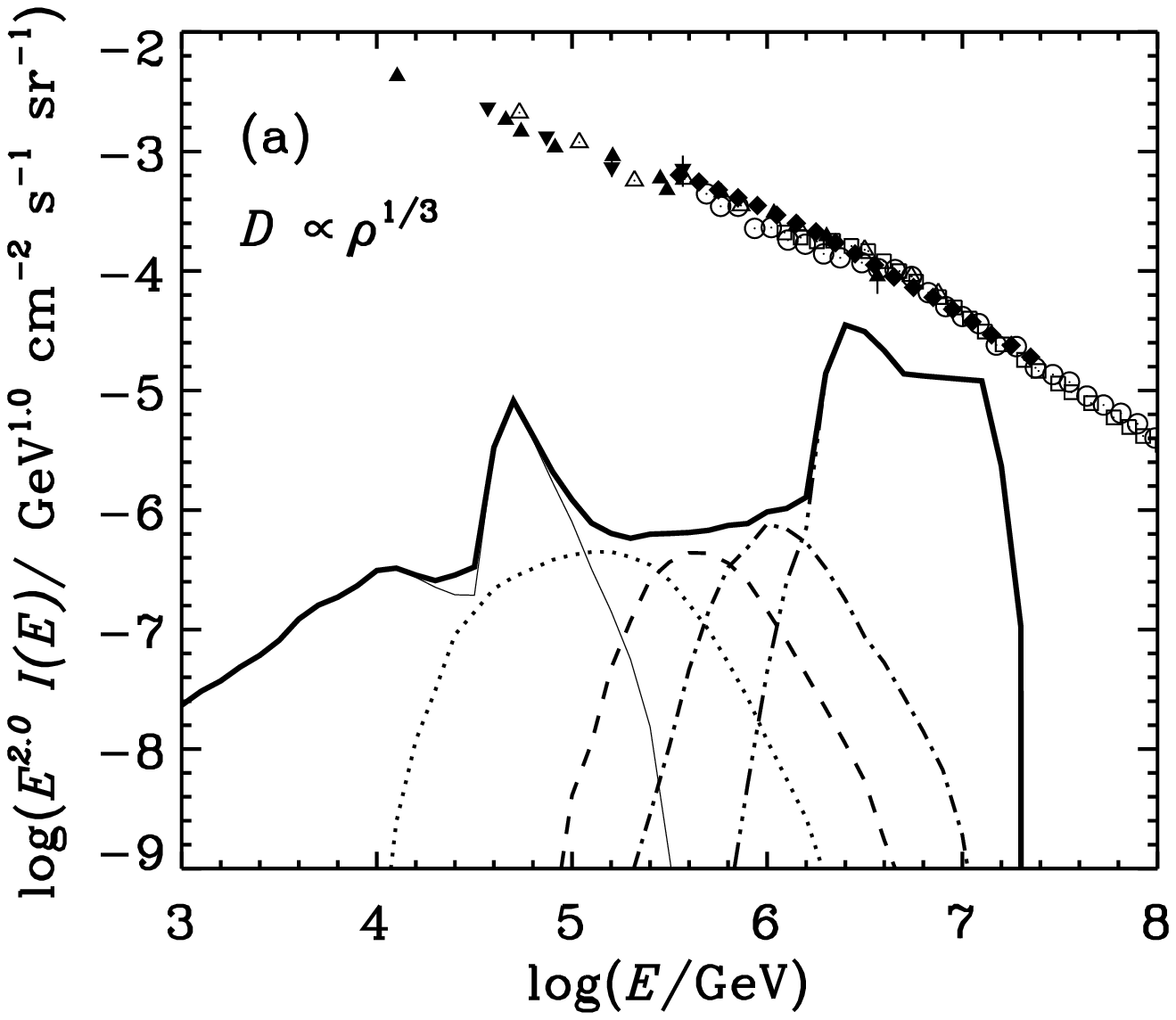, width=9cm}
\epsfig{file=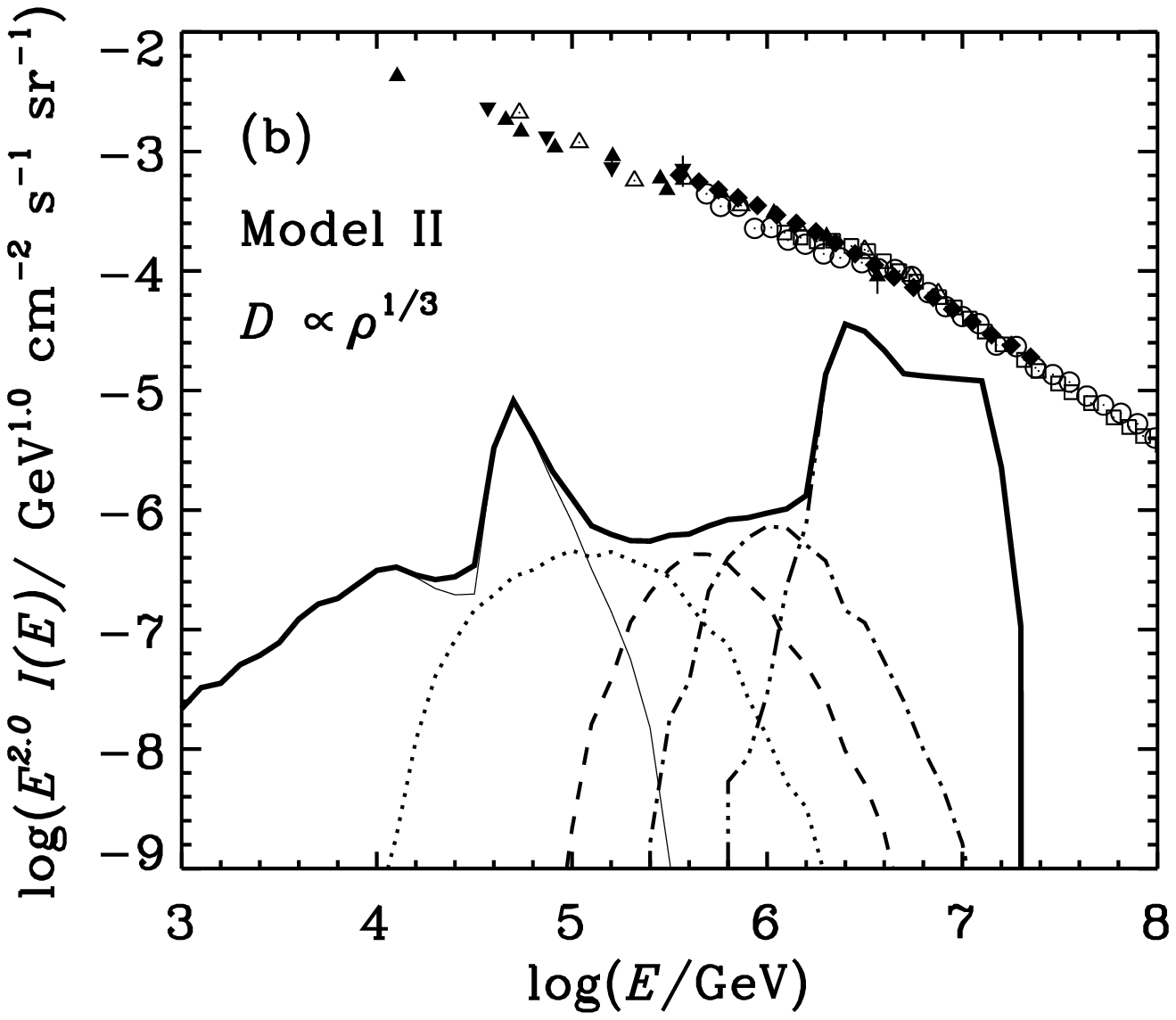, width=9cm}
\caption{Differential spectrum of nuclei injected by gamma-ray
pulsars after propagation as described in the text ($\xi=1$).  (a)
$D \propto \rho^{1/3}$; (b) $D \propto \rho^{0.54}$. Key to curves:
thick solid curve -- total; thin solid curve -- A=1; dotted curve
-- A=2--10; short dashed -- A=11--20; dot-dashed -- 21--40;
dot-dot-dot-dashed -- 41--56. The observed all-particle cosmic
ray spectrum spectrum taken from ref.~\protect\cite{GS98}.}
\label{fig:propspec0}
\end{figure}

\end{document}